# A MAGNETICALLY-TRIGGERED COMPOSITE MEMBRANE FOR ON-DEMAND DRUG DELIVERY

*Todd Hoare[1], Jesus Santamaria[2,3], Gerardo F. Goya[3], Silvia Irusta[2,3], Debora Lin[4], Samantha Lau[4], Robert Padera[5], Robert Langer[4], and Daniel S. Kohane[6]\**

[1]Department of Chemical Engineering, McMaster University, 1280 Main St. W, Hamilton, Ontario, Canada  L8S 4L7
[2]Networking Biomedical Research Center of Bioengineering, Biomaterials and Nanomedicine (CIBER-BBN). Zaragoza, Spain 50018
[3] Institute of Nanoscience of Aragón, University of Zaragoza, Pedro Cerbuna 12, Zaragoza, Spain 50009
[4]Department of Chemical Engineering, Massachusetts Institute of Technology, 45 Carleton St., Cambridge, MA, U.S.A. 02142
[5]Department of Pathology, Brigham and Women's Hospital, 75 Francis St., Boston, MA, 02115
[6]Laboratory  for Biomaterials and Drug Delivery, Department of Anesthesiology, Division of Critical Care Medicine, Children's Hospital Boston, Harvard Medical School, 300 Longwood Ave., Boston, MA , U.S.A. 02115

*To whom correspondence should be addressed
E-mail: Daniel.Kohane@childrens.harvard.edu

*Abstract*

Nanocomposite membranes based on thermosensitive, poly(N-isopropylacrylamide)-based nanogels and magnetite nanoparticles have been designed to achieve "on-demand" drug delivery upon the application of an oscillating magnetic field.  On-off release of sodium fluorescein over multiple magnetic cycles has been successfully demonstrated using prototype membrane-based devices.  The total drug dose delivered was directly proportional to the duration of the "on" pulse.  The membranes were non-cytotoxic, biocompatible, and retained their switchable flux properties after 45 days of subcutaneous implantation.

Delivery devices that allow remote, repeatable, and reliable switching of drug flux could have a marked impact on the treatment of a variety of medical conditions. An ideal device for on-demand drug delivery should safely contain a large quantity of drug, release little or no drug in the "off" state, be repeatedly switchable to the "on" state without mechanically disrupting the device, and be triggered non-invasively to release a consistent dosage demanded by a patient (e.g. local pain relief) or prescribed by a doctor (e.g. localized chemotherapy).

Despite the clear clinical need, few such drug delivery devices have been developed and none are available for clinical use. Existing technologies are particularly limited by their inability to be effectively triggered *in vivo* in the absence of a local implanted heat source, their lack of reproducible release over multiple thermal cycles, their slow response times to stimuli, and/or their inability to dynamically adjust drug dosing according to patient needs. Currently, no existing device overcomes all of these limitations. For example, radio frequency-activated microchips containing drug-filled reservoirs can achieve rapid on-demand drug delivery[1, 2] but deliver only fixed doses of drug and require implanted electronics. Near-IR responsive nanoparticles consisting of mixtures of PNIPAM and gold-gold sulfide nanoshells can release proteins on demand but deliver inconsistent doses upon multiple triggering cycles[3]. Ferrofluid-loaded polymer sheets[4], liposomes[5], microspheres[6, 7], microcapsules[8, 9], and nanospheres[10-12] can be activated remotely by magnetic induction *but typically achieve either single burst release events or inconsistent dosing over multiple thermal cycle*s due to the use of mechanical disruption of the drug-polymer matrix as the flux triggering mechanism. Hence, alternative technologies are needed.

Hydrogels[13-18], gel-based microparticles[19] or nanoparticles[20-23], and surface-grafted polymers[24-36] based on thermosensitive poly(N-isopropylacrylamide) (PNIPAM) have been frequently used in triggerable devices. With heating, PNIPAM undergoes a reversible discontinuous phase transition in water, switching from hydrophilic to hydrophobic[37]. In a PNIPAM-based hydrogel, this phase transition induces a deswelling response which typically reduces drug flux from the hydrogel. Alternately, when PNIPAM is used to fill the pores of a membrane, the pores are opened upon heating as the entrapped polymer shrinks, increasing drug flux through the membrane[28, 38]. Such membranes have been designed by grafting poly(N-isopropylacrylamide) to existing membrane networks[5] or by entrapping PNIPAM microgels within a membrane matrix[39]. However, existing PNIPAM-based devices would be permanently "on" at physiological temperature (37°C) since their transition temperatures are ~32°C.

Existing technologies would also require use of an implanted heating system for effective *in vivo* activation.

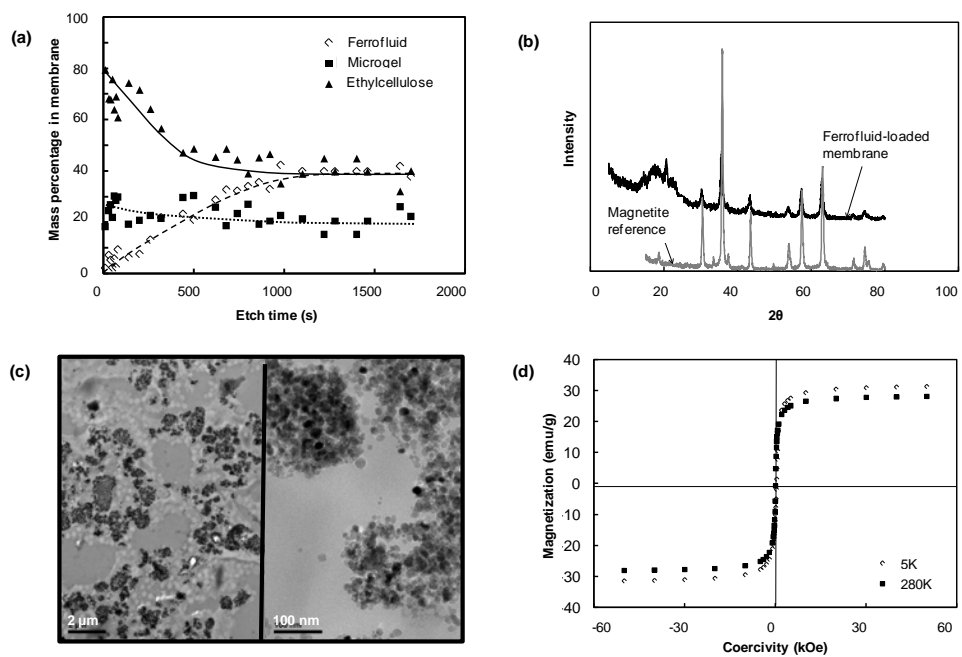

*Figure 1.* *Physicochemical membrane characterization: (a) Mass percentages of ferrofluid, nanogel, and ethyl cellulose in membrane as a function of etch time, by XPS; (b) XRD spectrum of ferrofluid-loaded membranes in comparison to a magnetite-only control; (c) Transmission electron micrographs of ferrofluid distribution and size within the composite membrane; (2 μm size bar, left panel; 100nm size bar, right panel); (d) Magnetization curves for composite membranes measured at 5K and 280K.*

Here, we developed a composite membrane based on multiple engineered smart nanoparticles which enabled rapid, repeatable, and tunable drug delivery upon the application of an external oscillating magnetic field. The membrane consisted of ethylcellulose (the membrane support), superparamagnetic magnetite nanoparticles (the triggering entity), and thermosensitive poly(N-isopropylacrylamide) (PNIPAM)-based nanogels[37] (the switching entity). Membranes were prepared by co-evaporation so that the nanogel and magnetite nanoparticles were entrapped in ethylcellulose to form a presumably disordered network. Surface-etching x-ray photoelectron spectroscopy (XPS) showed that the membranes had a relatively uniform composition within the bulk but relatively less iron (ferrofluid) near the membrane surface (Figure 1a). The membrane nanogel composition determined by XPS (23% by dry weight) correlated well with the nanogel concentration in the pre-membrane suspension (25% by dry weight).

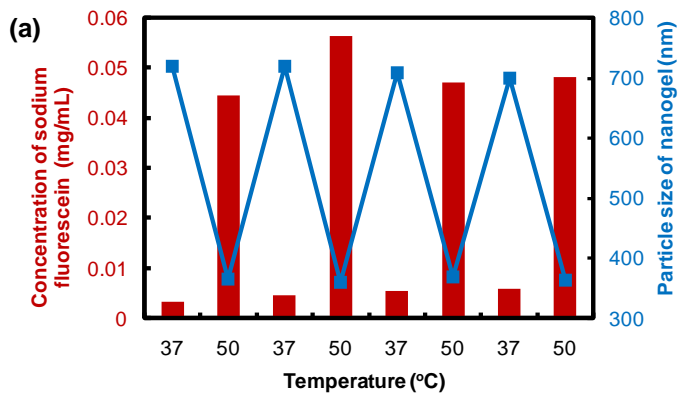

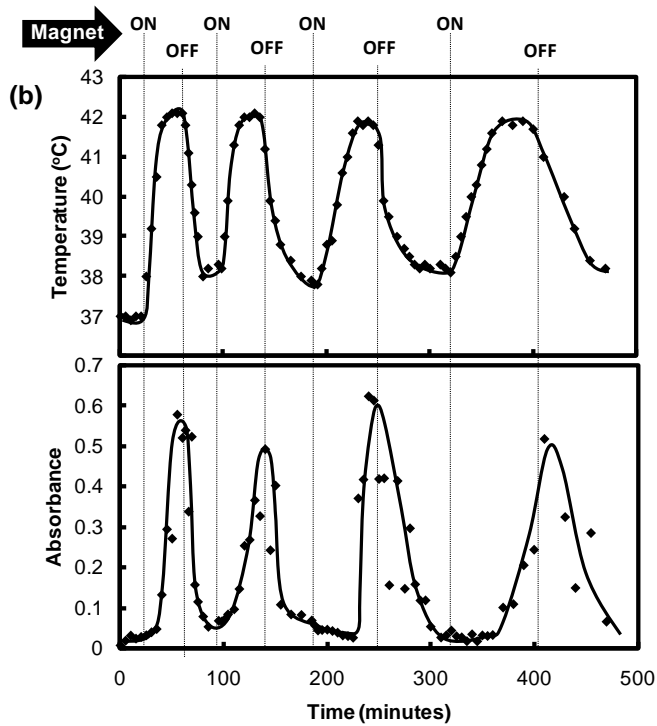

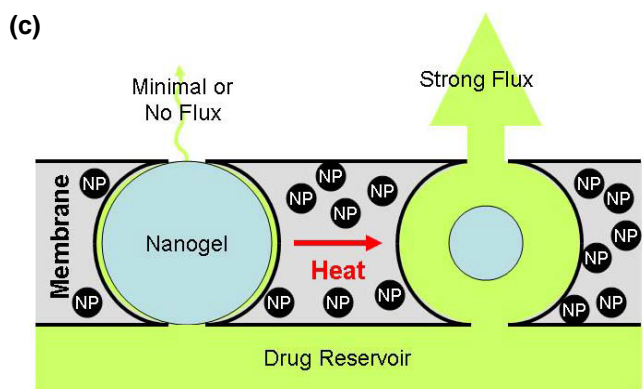

*Figure 2.* Stimulus-responsive membrane triggering in vitro: (a) Temperature-triggering: comparison of nanogel particle size in suspension (blue data, right y-axis) and differential flux of sodium fluorescein through the nanogel-loaded membranes (red data, left y-axis) as a function of temperature; (b) Magnetic triggering: temperature profile in the sample chamber and differential flux of sodium fluorescein out of membrane-capped devices as a function of time over four successive on/off cycles of the external magnetic field; (c) schema of the proposed mechanism of membrane function

X-ray diffraction analysis suggested that ferrofluid particles had a magnetite crystal structure and an average crystallite size of ~12 nm (Figure 1b). Transmission electron microscopy of a membrane section (Figure 1c) confirmed the average ferrofluid particle size of 10-25 nm and suggested *that local ferrofluid clusters sized between 0.1 - 3 μm were distributed throughout the membrane*. The magnetic material within the membranes had a magnetic saturation value of 96.5 emu/g($Fe_3O_4$) at 280 K (Figure 1d), similar to values previously reported for bulk magnetite (93-96 emu/g)[40]. Furthermore, the measured coercive field of 346 ± 4 Oe at 5 K is consistent with that of previously reported ferrofluid particles of similar size[41]. These results suggested that the ferrofluid particles consisted of a single magnetic domain (i.e. all iron was in magnetite form) and had the superparamagnetic properties and average particle size required for effective magnetic induction heating in an oscillating magnetic field[42].

To facilitate effective *in vivo* triggering, the nanogels were engineered to remain swollen (i.e. in the "off" state) at physiological temperature by copolymerizing N-isopropylacrylamide (NIPAM) with N-isopropylmethacrylamide (NIPMAM) and acrylamide (AAm). The methyl group of NIPMAM sterically inhibits the phase transition[43] while AAm is more hydrophilic than NIPAM[44], both shifting the phase transition to higher temperatures. The ratio between the monomers was chosen to maximize the size change from the swollen to the collapsed state, in order to optimize membrane pore opening when triggered.

The ability of the membrane constituents and the composite membrane to trigger at physiologically relevant temperatures was evaluated using both thermal and magnetic stimuli. Nanogels in free suspension in PBS underwent a ~400 nm change in diameter upon heating from physiological temperature to 50°C (Figure 2a), with >90% of the total deswelling transition completed at 43°C. Thermal triggering of the nanogel-containing membrane was tested by placing it between two chambers of a glass flow cell submerged in a water bath and evaluating the flux of sodium fluorescein across the membrane (i.e. between the chambers) as a function of time and temperature. A ~20-fold higher flux of sodium fluorescein occurred at temperatures exceeding the volume phase transition temperature (~40°C) of the nanogels (Figure 2a). FT-IR analysis confirmed that this permeability enhancement coincided with a change in the hydrogen bonding within the membrane, consistent with the occurrence of a nanogel volume phase transition. Furthermore, the fluorescein flux could be switched on and off over multiple thermal cycles with high reproducibility, suggesting that the nanogel phase transition inside the membrane pores was fully reversible.

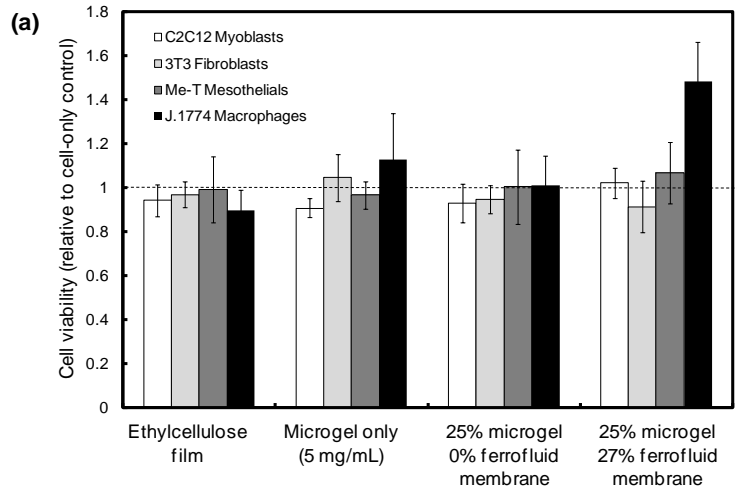

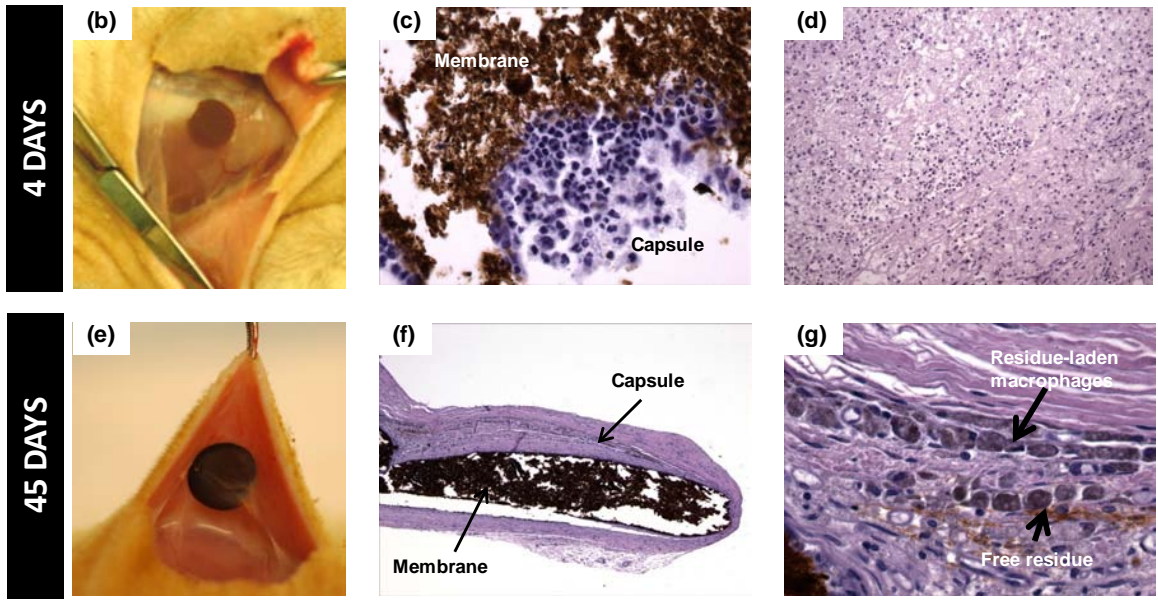

*Figure 3. Biological testing of membranes: (a) Cell viability (relative to a cell-only control well) for differentiated myoblasts, fibroblasts, mesothelial cells, and macrophages in the presence of membrane components and membranes; (b-g) Tissue response to implanted nanogel- loaded membrane (25% nanogel, 27% ferrofluid) after 4 and 45 days of implantation: (b) top view, 4 days post-implantation; (c) histological section of membrane-tissue interface, 400x magnification; (d) histological section of capsule inflammatory response, 100x magnification; (e) top view, 45 days post-implantation; (f) histological section of membrane-tissue interface, 40x magnification; (g) histological section of capsule inflammatory response, 400x magnification.*

Magnetic triggering was evaluated in small-scale devices made by gluing two 1 cm diameter membrane disks to the ends of a 1 cm length of silicone tubing filled with a sodium fluorescein solution. The devices were mounted singly inside a semi-adiabatic flow cell in a solenoid coil, with constant water flow through the flow cell to permit continuous sampling of fluorescein release. Figure 2b shows the magnetic triggering of the composite membrane. The magnetic nanoparticles embedded in the membrane heated inductively when subjected to an external oscillating magnetic field, heating previously attributed to power absorption and subsequent magnetic relaxation of single-domain nanoparticles[45]. At the applied magnetic frequency and field amplitude, the water inside the semi-adiabatic flow cell heated from 37°C to ~42°C over the course of ~10 minutes, at which point the temperature reached steady state. Heat generated by magnetite induction heating was transferred to the adjacent thermosensitive nanogels, causing the nanogels to shrink and permit drug diffusion out of the device. When the magnetic field was turned off, the device cooled, causing the nanogels to re-swell and refill the membrane pores. As a result, the drug flux returned back to a near-zero value (Figure 2c). As in the thermally-activated experiments, a 10-to-20-fold differential flux was observed between the "off" and "on" states. Furthermore, multiple on-off cycles could be performed without significantly changing the permeability of the membrane in the "off" state. This reproducibility suggests that magnetically-triggered physical distortion of the device[42] plays no significant role in accelerating drug release from the membrane-based devices.

| Cycle | Duration of "on" cycle (minutes) | Total mass released (mg) | Rate of drug release (mg/min) |
|---|---|---|---|
| 1 | 35 | 0.43 | 0.012 |
| 2 | 40 | 0.47 | 0.012 |
| 3 | 57 | 0.69 | 0.012 |
| 4 | 75 | 0.83 | 0.011 |

**Table 1.** Total mass of sodium fluorescein release and rate of drug release during each magnetic cycle shown in Figure 2b

The membrane-based devices also permitted precise control of the amount of drug released as a function of the duration of the magnetic pulse. Table 1 shows the dose of fluorescein delivered for each of the four magnetically-activated cycles shown in Figure 2, calculated by integrating the area under the absorbance vs. time curve for each cycle. The mass of compound released over each triggering cycle varied directly with the duration of the magnetic pulse ($R^2 = 0.995$), with the rate of drug release varying by less than 10% in each cycle. Thus, drug release could be controlled by modulating both the frequency and duration of magnetic pulse.

The devices turned "on" with only a 1-2 minute time lag after the solution temperature reached 40°C and turn "off" with a ~5-10 minute lag from the cooling temperature profile (Figure 2b). This response rate was much more rapid than that seen with bulk, interpenetrating hydrogel networks, which can exhibit swelling kinetics on the order of hours[46].

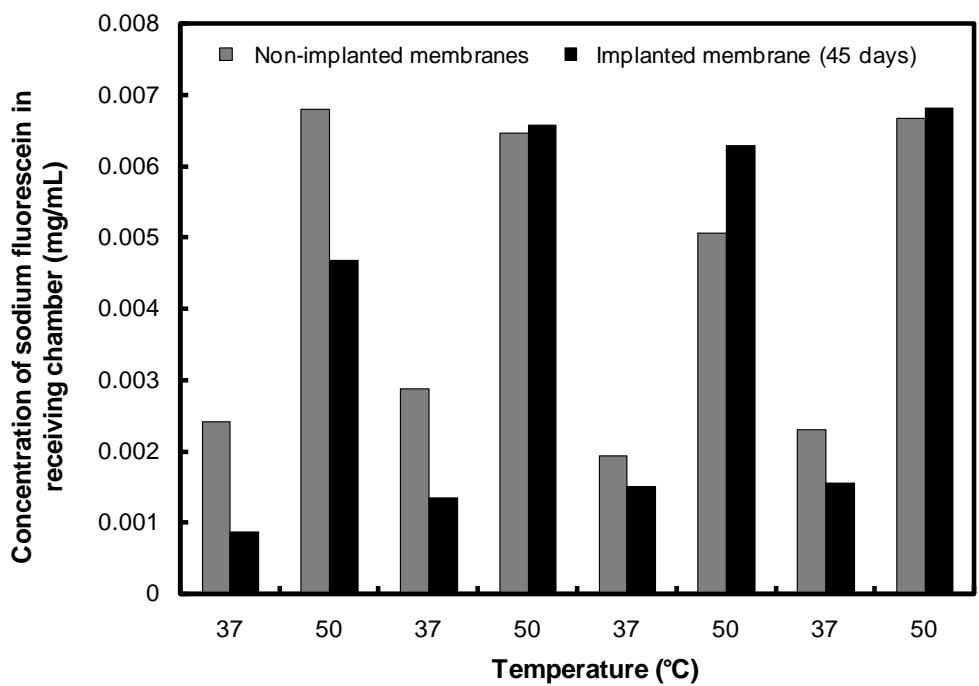

*Figure 4. Comparison of thermally-triggered membrane flux of a freshly-prepared membrane to a membrane explanted from a Sprague-Dawley rat after 45 days of implantation.*

To evaluate potential applicability of the membranes for *in vivo* drug delivery, we first evaluated the cytotoxicity of the membranes to a broad range of cell types (differentiated myoblasts, fibroblasts, macrophages, fibroblasts, macrophages, and mesothelial cells). Figure 3a shows the viability of cells in media exposed to a composite membrane or its components (an ethylcellulose film, a 5 mg/mL copolymer nanogel suspension, and a nanogel-loaded membrane), expressed as a ratio to cell survival in media alone as measured by the MTT assay. No significant decrease in cell viability was observed in any cell line upon exposure to the composite membrane or its individual components.

Biocompatibility of 1 cm diameter composite membrane disks was tested by subcutaneously implanting 1 cm diameter membrane disks in Sprague-Dawley rats. Rats were sacrificed at predetermined intervals, at which point the membrane and surrounding tissues were removed and analyzed by histology. Representative tissue responses at 4 and 45 days post-implantation are shown in Figures 3b-g. After 4 days, the membrane was not significantly walled off from the surrounding tissues, with only minimal tissue adhering to the membrane (Figure 3b). The gross appearance of the implant at 4 days was bland with only mild erythema consistent with the recent implantation. Microscopically, there was acute and early chronic inflammation around the implant, as would be expected at this time point (Figures 3c and 3d). At 45 days, there was a thin translucent tissue capsule around the implant (easily separable from the membrane by gentle dissection) with no evidence of tissue damage (Figures 3e and 3f). The implants were grossly intact at both time points. The sections showed a mature fibrous capsule with macrophages and occasional foreign body giant cells at the material-tissue interface (Figure 3g). Occasional macrophages containing implant material were present in the tissue capsule, along with some residual free membrane material (Figure 3g). There was no apparent amplification of an inflammatory response and no evidence of ongoing acute inflammation.

To assess whether membranes retained their inducible drug-releasing properties *in vivo*, a membrane was excised after 45 days of subcutaneous implantation, the thin tissue capsule was removed, and the thermally-triggered fluorescein flux was measured using a glass flow cell apparatus. The flux response of the excised membrane was compared to that of a fresh, non-implanted membrane with the same composition, as shown in Figure 4. No significant difference was observed in the flux differential between the "on" and "off" states or the absolute magnitude of fluorescein flux across the membrane before or after implantation. This result suggests that protein adsorption or biofilm formation *in vivo* does not significantly impact the functionality of the membrane.

The composite nanogel-ferrofluid membrane described here meets the important criteria for "on-demand" drug delivery devices. It can undergo a sharp, discontinuous volume phase transition at ≥40°C and so can be switched from the "off" state at normal physiological temperature to the "on" state at a temperature where it would not typically be triggered by fever, local inflammation, etc. The membrane could be switched on and off rapidly by the application and removal of an external oscillating magnetic field. Thus, on-demand drug delivery could be triggered non-invasively without implanted electronics. Furthermore, the membrane remained stable during multiple magnetic triggering cycles and over extended *in vivo* implantation, making reproducible, multi-cycle drug delivery possible. In each case, the functionality of the membrane was directly attributable to the nanoparticle properties; specifically, the rapid swelling kinetics and engineered phase transition behavior of the nanogel and the surface chemistry and optimized size of the magnetite nanoparticles for magnetic induction heating enabled the rapid, repeatable, and tunable drug release properties observed under physiological conditions.

Composite membrane-based drug delivery devices have the potential to greatly increase the flexibility of pharmacotherapy and improve the quality of patients' lives by providing repeated, long-term, on-demand drug delivery for a variety of medical applications, including the treatment of pain (local or systemic anesthetic delivery), local chemotherapy, and insulin delivery. Modulation of the magnetic field could allow for fine-tuning of the rate of drug release, in addition to the frequency and duration of treatments. Additionally, the ability of the membranes to remotely and reversibly control chemical permeation may be applied in the design of triggered bioseparation modules, selective chemosensors, or externally activated microreactors.

**Acknowledgements:** This research was funded by NIH grant GM073626 to DSK. TH acknowledges post-doctoral funding from the Natural Sciences and Engineering Research Council of Canada. SI and GFG acknowledge support from the Spanish MEC through the Ramon y Cajal program.

**Supporting Information Available:** Descriptions of the materials and methods used, extensive physicochemical characterization of the membranes, thermal phase transition profiles of the nanogels, and drug release kinetics from the composite membranes are all reported in the supporting information. This material is available free of charge via the Internet at http://pubs.acs.org.